# Near dispersion-less surface plasmon polariton resonances at a metal-dielectric interface with patterned dielectric on top

Sachin Kasture[1], P. Mandal[1a)], Amandev Singh[1], Andrew Ramsay[2], A. S. Vengurlekar[1], S. Dutta Gupta[3], Vladimir Belotelov[4], and Achanta Venu Gopal[1b)]

[1]DCMPMS, Tata Institute of Fundamental Research, Homi Bhabha Road, Mumbai-400005, India

[2]Department of Physics and Astronomy, University of Sheffield, Sheffield S37RH

[3]School of Physics, University of Hyderabad, Hyderabad 500046

[4]M.V. Lomonosov Moscow State University, 119991 Moscow, Russia

**Abstract:** We report possibility of SPP modes that can be excited by incident light of a fixed frequency coming at wide angles. We experimentally show such modes in structures with smooth dielectric-metal-dielectric interfaces having 2-D dielectric patterns on top. Calculated field profile establishes the field localization at the metal-dielectric interfaces. We show that the position and dispersion of the excited modes can be controlled by the excitation geometry and the 2-D pattern.

a) Currently at IIT, Kanpur
b) Email id: achanta@tifr.res.in

# 1. Introduction

Surface modes at an interface especially surface plasmon polaritons (SPPs) at metal-dielectric interface are well known solutions of Maxwell's equations subject to appropriate boundary conditions. Such modes are typically excited using prism (in Otto, Kretschmann or Sarid configurations) or grating coupling.[1-3] For waveguide applications, dielectric loaded SPPs are also widely studied.[4] In the context of grating coupling, several structures like periodically patterned dielectric with metal deposited on top, patterned metal structures, metal nanoparticle arrays on dielectric or dielectric pattern on unpatterned metal-dielectric interfaces have been studied. Both basic studies related to understanding plasmon dispersion and applications involving plasmon mediation have been reported in these structures.[5-17] Coupling of plasmon modes to waveguide modes has also been studied.[18-19] Inspite of the vast literature, one of the structures that is less studied is a 2D dielectric pattern on dielectric-metal-dielectric heterolayer structure. In this paper we study 3 layer structure with 2D dielectric pattern on top and present possibility of modes that can be excited by light of given frequency incident at wide angles.

These modes are thus different from the dispersion-less (flat) localized or particle plasmon and the waveguide modes arising in unpatterned dielectric. Previously, $\theta$ independent SPP dispersion was shown in 1D metal gratings for TE polarized light with azimuthal angle of $90^0$.[20] The present angle independent SPP modes are excited by TM polarized light and can be controlled by the measurement geometry as will be shown below.

The structure of this letter is as follows: in section II we give a theoretical description of the conditions under which launch angle independent excitation of SPP modes are possible in 2D grating structures. In section III we present the sample preparation followed in Section IV by the experimental results on 2D dielectric pattern on dielectric-metal-dielectric heterolayer structures and numerical simulation results for the field profile showing field confinement at the interfaces. In section V we summarize the results.

**2. Origin of the flat mode**

In this section, we will first describe the geometry followed by the conditions under which the launch angle independent SPP modes are possible. The measurement geometry is shown in Fig.1 with a 2D periodic pattern in x-z plane with lattice constants in x and z-directions given by $a_1$ and $a_2$, respectively. For light propagation along y-direction, launch angle, $\theta$ is the angle subtended by light wave vector k with respect to the normal to the interface (y-axis) and $\varphi$ is the azimuthal angle in the xz-plane. Thus, for TM polarized light propagating along y-direction the relevant electric field component, $E_x$, is along $a_1$ for $\varphi = 0°$.

For a single metal-dielectric interface with a 2D periodic structure on top, light is coupled to plasmon modes matching one of the Bragg harmonics of the incident light energy. In addition, the momentum conservation requires matching of the light momentum k= ($k_0 \sin\theta\cos\varphi$, $k_0\cos\theta$, $k_0\sin\theta\sin\varphi$) and the wave vectors $k_{G,a}$ = ($\pm k_{x,G,ax}$, $k_y$, $\pm k_{z,G,az}$) where, G is the grating vector. For a 2-D system with $2\pi/a_1$ ($G_x$) and $2\pi/a_2$ ($G_z$) as grating vectors in reciprocal space in the two orthogonal directions, and m and n are integers, the SPP dispersion relation is given by Eq. (1) below.[21]

$$\sqrt{(k_0 \sin(\theta)\cos(\varphi) \pm mG_x)^2 + (k_0 \sin(\theta)\sin(\varphi) \pm nG_z)^2} = k_0\sqrt{\varepsilon'} = k_{SPP} \quad (1)$$

where, $\varepsilon' = \varepsilon_d\varepsilon_m/(\varepsilon_d+\varepsilon_m)$ in which $\varepsilon_d$ and $\varepsilon_m$ are dielectric constants of dielectric and metal layers that are wavelength dependent. In Eq. (1), we consider the real part of the dielectric constants. $k_0=2\pi/\lambda_0$ where $\lambda_0$ is the resonant wavelength. Eq. (1) shows that for given interface (that is, $\varepsilon'$), each resonant $k_0$ corresponds to a $k_{SPP}$ based on the launch geometry $(\theta,\varphi)$ and the 2D pattern ($a_1$ and $a_2$) for given m and n.

For a given wavelength (and thus $k_0$), $\varepsilon'$ is constant and the solutions to $k_0$ in Eq. (1) are given by the following for $A=2\sqrt{2}\pi\sin\theta$, $M=m\cos\varphi/a_x$ and $N=n\sin\varphi/a_z$,

$$k_0 = \frac{[-A(M+N)] \pm \sqrt{[A(M+N)]^2 - 4(\sin^2\theta - \varepsilon')(G_x^2 + G_z^2)}}{2(\sin^2\theta - \varepsilon')} \quad (2)$$

From Eq. (2), thus, for a given wavelength (or mode), the launch angle ($\theta$) dependence is coming from the terms ($\sin^2(\theta)-\varepsilon'$) and A. The limitation set by these two terms can be overcome by the choice of materials (for metal-dielectric interface, $|\varepsilon'| >> \sin^2(\theta)$) and the geometry (to make $A(M+N)=0$). The three cases under which $A(M+N)=0$ for non-zero $\theta$ are given by (i) $\varphi = 45°$ and $m = -n$, or (ii) $\varphi = 0°$, $m = 0$ or (iii) $\varphi = 90°$, $n = 0$. For case (i), the weak $\theta$-dependence is possible only in one half of the dispersion plane (either positive or negative $\theta$). Experimental demonstration of such modes can be seen in Ref.11. In principle, for case (ii) (and case (iii)), weak $\theta$-dependence is possible for all n (and m) values. While for square lattice, due to symmetry, cases (ii) and (iii) are degenerate, for non-square lattice this degeneracy is lifted and different near dispersion-less modes can be excited based on the orientation.

In order to demonstrate these θ-independent modes experimentally, we choose dielectric-metal-dielectric layer structure with 2D pattern on top. For unpatterned three layer structure with a thin metal layer sandwiched between two infinite layers, the general dispersion relation for surface modes is given in terms of the dielectric constants ($\varepsilon_1$, $\varepsilon_2$, and $\varepsilon_3$) and thickness of 2$^{nd}$ layer (t) by,[22]

$$\tanh(\alpha_2 t) = -\frac{\varepsilon_2 \alpha_2 (\varepsilon_1 \alpha_3 + \varepsilon_3 \alpha_1)}{(\varepsilon_1 \varepsilon_3 \alpha_2^2 + \varepsilon_2^2 \alpha_1 \alpha_3)} \qquad (3)$$

where $\alpha_j^2 = k^2 - k_0^2 \varepsilon_j$, j = 1, 2, 3 with k being the propagation constant along the interface. Eq. (3) gives the coupled SPP mode solutions where the coupling strength depends on the thickness of the gold film (t). In order to calculate the SPP dispersion in three layer structure with 2D pattern on top, we substitute the $k_{SPP}$ value from Eq. (1) in Eq. (3). In the following we experimentally demonstrate the launch angle independent, near flat dispersion modes in non-square and square lattices.

## 3. Sample preparation

Two dimensional (2d) plasmonic crystals with arrays of air holes in dielectric have been fabricated by interference lithography. On fused quartz pieces, thin gold (Au) layer was deposited by thermal evaporation followed by the spin coating of 500nm thick photoresist (SHIPLEY S1805). Subsequently 2-D lattice of holes or pillars were obtained by double exposure of the sample to interference pattern of a laser beam (442 nm He-Cd line) followed by developing in standard developer.

The grating parameters of two of the structures estimated from atomic force microscopy are given below. Rectangular lattice sample has $a_1$= 730nm and $a_2$ = 660nm, thickness of the gold layer (t) is 55nm, depth of the air holes (d) is 60nm and fill factor (fraction of total period occupied by air) is 0.4 in X-direction and 0.45 in Z-direction. Thus, the diameter of air hole in X-direction is 292nm (fill factor x period). SEM image of the top pattern is shown in Fig. 1(b). Relevant parameters for the near square lattice sample are 780nm, 750nm, 40nm, 80nm, 0.61, and 0.68, respectively. Each of the samples has a 400nm thick (h) unpatterned photoresist (dielectric) above the gold layer.

## 4. Experimental results and discussion

SPP dispersion in different samples was studied by angle resolved white light transmission measurements with a collimated 100W tungsten halogen lamp output and a fiber based spectrometer having 0.5nm wavelength resolution in the 400-1000nm wavelength range. A Glann-Thompson polarizer (extinction ratio $10^5$:1) is used for linearly polarizing incident light to TM. The spectra are normalized with those through an unpatterned reference sample.

SPP dispersion (wavelength vs angle of incidence) is plotted in Fig. 2 for rectangular lattice taken between -25º and 25º incident angles at 0.3º steps. Spectra are presented in a contour plot with transmission dips shown to be dark. Thus, the darker curves in the contour plot represent dips in the transmission spectrum.

The measured SPP dispersion curves are fitted to Eq (3) with the solutions for in-plane momentum taken from the left hand side of Eq. (1). The fits are presented as white dashed lines in Figures. The real part of the dielectric constant of gold is used for the fits where the values are taken from Johnson-Christie.[23] The best fits are for dielectric constant of the resist and quartz to be 2.7 and 2.15, respectively. The surface roughness of each of the layers (less than 1nm), thickness of gold and the grating parameters are measured using profilometer and AFM.

The orientation of the sample is varied from $\varphi \sim 0°$ to $\varphi \sim 90°$. In Fig. 2, SPP dispersion at two orientations, $\varphi \sim 0°$ (a) and $\varphi \sim 90°$ (b) are shown for rectangular lattice. Both Rayleigh anomalies (obtained by equating the left hand side of Eq. (1) to $k_0\sqrt{\varepsilon_d}$) labeled $(m,n)_G$ and the SPP modes labeled $(m,n)_P$ (obtained by substituting left hand side of Eq. (1) and solving Eq. (3)) are observed.

The near dispersion-less mode $(0, 2)_P$ at about 650nm seen in this sample belongs to case (ii) discussed above. For $\varphi \sim 0°$ the sample is oriented such that the 660nm period is along z-axis and thus the flat mode near this wavelength is seen. For $\varphi \sim 90°$, the flat mode should be at about 730nm and is seen accordingly in Fig. 2(b). One can thus control the SPP dispersion by controlling the orientation ($\varphi$).

Due to the coupled modes (in the direction normal to the interface) at the dielectric-metal and metal-quartz interfaces, we expect the modes to split into symmetric and anti-symmetric modes. Some of the observed coupled modes are labeled in the figure like $(\pm 2,1)_A$ and $(\pm 2,1)_S$, where subscript A stands for Antisymmetric and S for Symmetric for given (m,n) mode.

To further establish the field localization due to plasmon excitation at the two interfaces, we performed finite difference time domain (FDTD) simulations on the 3 layer structure with 2D dielectric pattern on top. In order to calculate the field profile at the top and bottom interfaces (X or Z cut) and in-plane (Y-cut) FDTD simulations are carried out using Lumerical's FDTD Solutions. The structure and material parameters are as mentioned above. With a period of 660nm, using periodic boundary conditions in the X- and Z-directions and perfectly matched layers in the Y-direction (PML reflection of 1e-9) with 6nm step in each direction, simulations are done over a unit cell of one period. At the metal-dielectric interface a finer mesh of 4nm is used.

We used a plane wave covering 400-800nm wavelength range centered at about 600 nm to calculate the field at different wavelengths. The normalized $|E|^2$ (with respect to the unpatterned structure) clearly shows (Fig. 3) the field localization at the top and bottom interfaces for resonant wavelength of 680nm. Away from the resonant wavelength, field localization at the interfaces goes to zero.

Fig. 4 shows the contour plot of the measured SPP dispersion for square lattice of air holes in dielectric with the incident light polarization set to TM polarized light for $\varphi \sim 0°$. Also shown are the fits to the data based on model presented above. Data and calculations show multiple SPP modes including the near dispersion-less mode corresponding to $(0,2)_P$ at about 805nm. With the material parameters given above, there is good match between the measured and the calculated dispersion. These results show that, in heterolayers, it is possible to excite SPPs by placing a patterned thin film on top as a mask to modulate the incoming light. Possibility of the launch angle independent modes shows that these would be useful for applications such as in photovoltaics and plasmon mediated enhancement of optical properties at specific frequencies.

## 5. Conclusion

The signature of a sub-wavelength two-dimensional dielectric grating layer on coupled surface plasmon polariton modes in a layered configuration is studied. The grating layer comprises of square or rectangular arrays of sub-wavelength air holes in the dielectric. Near flat, launch angle independent, SPP dispersion curves are shown. Controlling the top pattern and launch conditions would help control the position of the near flat mode. This could have wide practical applications in omni-directional light coupling to plasmon modes to make use of plasmon mediated near-field enhancement.

**ACKNOWLEDGMENTS**

Partial financial support for this work is provided by Department of Science and Technology (DST,India)-UK-India Education and Research Initiative (UKIERI).

Figure Captions:

Fig.1 (a) Schematic of the dielectric-metal-dielectric heterolayer with 2D top pattern in the X-Z plane is shown. Launch angle (θ), of incident light of wave vector k, is shown with respect to the normal to the interface (y-axis). φ is the azimuthal angle such that 2-D pattern has lattice constant $a_1$ along X- and $a_2$ along Z-direction for φ=0°. Depth of the pattern (d), height of top dielectric (h) and thickness of Gold (t) are also shown. SEM image of the top pattern (rectangular lattice) with the sample orientation for φ=0° and φ≠0° is shown in (b).

Fig. 2: Measured SPP dispersion is shown for rectangular lattice of air holes in dielectric on top of smooth dielectric-metal-dielectric structure with φ ~ 0° (a) and φ ~ 90° (b). Scale bar of the contour plots were shown on the top of each plot. White dashed lines are fits. Mode descriptions are given in text.

Fig.3 Calculated intensity profile for φ=0° and resonant wavelength of 680 nm for the three layer structure with top dielectric having 2D lattice of air holes with period 660nm and air hole depth of 60nm. White line at the top shows the air hole position and the dark thin region vertically centered at about 470nm is the Gold film with quartz underneath and unpatterned resist on the top.

Fig.4 Calculated (dashed lines) and measured SPP dispersion for a square array of air holes in resist on top of dielectric-metal-dielectric layer structure for φ~0°. Launch angle independent mode at about 805nm is seen in both.

Figure 1 S. Kasture etal

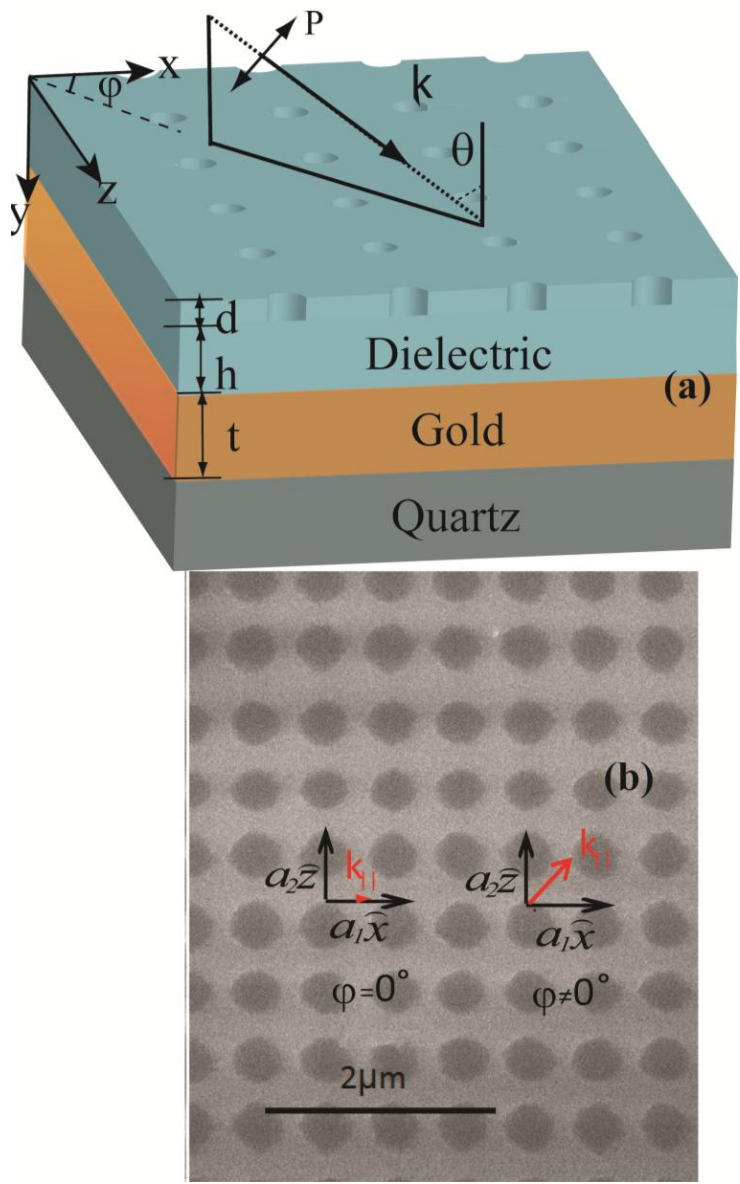

Figure 2  S. Kasture etal

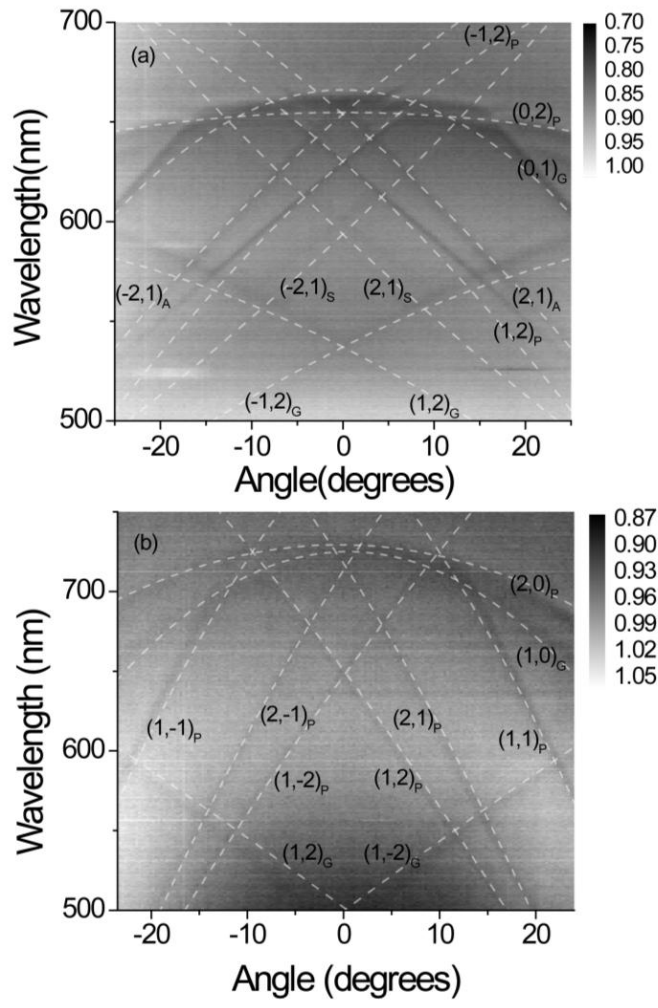

Figure 3 S. Kasture etal

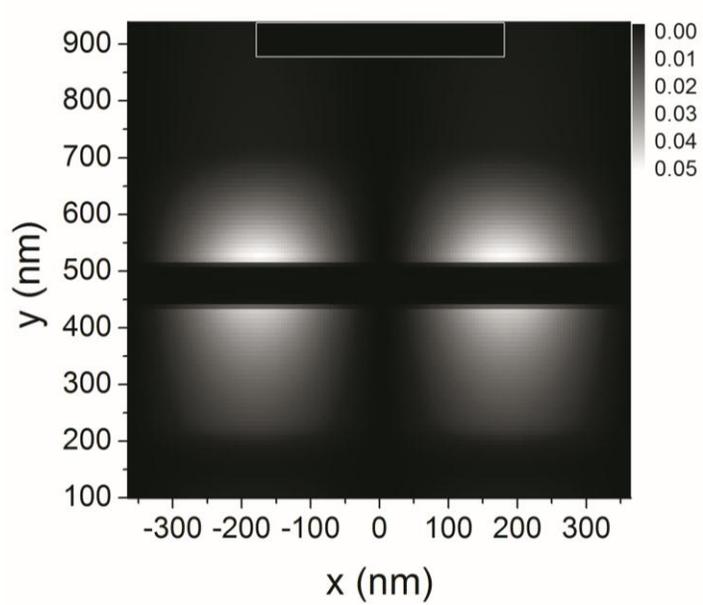

Figure 4 S. Kasture etal

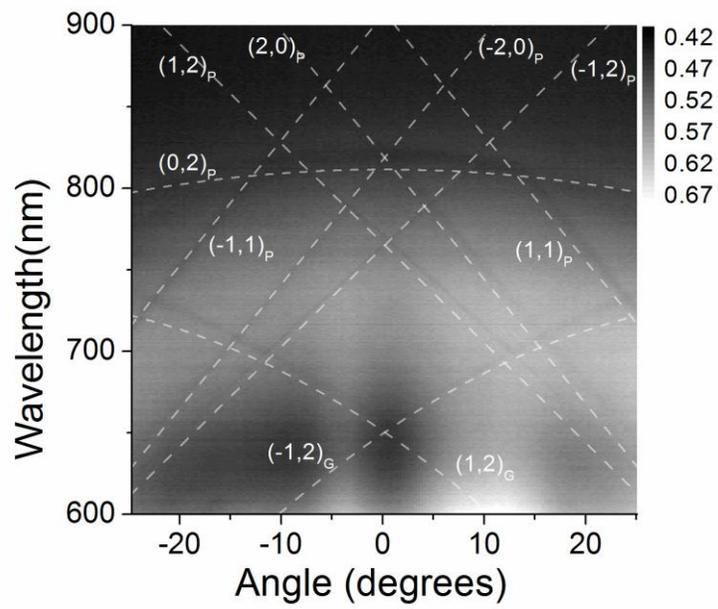